\documentclass[conference]{IEEEtran}
\usepackage{blindtext, graphicx}
\usepackage{array}
\usepackage{booktabs}
\usepackage{amsmath}
\usepackage{footnote}
\usepackage{hyperref,cite}

\begin{document}

\title{Social Data Analysis: A Study on Friend Rating Influence}

\author{
\IEEEauthorblockN{Shubhanshu Gupta}
\IEEEauthorblockA{Truebil\\
Mumbai, India\\
Email: shubhanshu.gupta93@gmail.com}\\ 
\and
\IEEEauthorblockN{Vaibhavi Desai}
\IEEEauthorblockA{Google India, Bangalore\\
Email: rani.vhd@gmail.com}\\[0.9cm] 
\and
\IEEEauthorblockN{Harsh Thakkar}
\IEEEauthorblockA{Enterprise Information Systems Lab,\\
University of Bonn\\
Bonn, Germany\\
Email: hthakkar@uni-bonn.de}\\
}

\maketitle

\begin{abstract}
%\boldmath
Social Networking accounts for a significant chunk of interest among various online activities~\cite{smith2009social}. The proclivity of being social, online, has been ingrained in us so much that we are actively producing content for the rest of the world to see or take interest in our whereabouts, our meals, our opinions, photographs etc. Yelp\footnote{Yelp~(\url{https://www.yelp.com/})}, seamlessly, integrates this very aspect of people in its portal. It engages people to write reviews about the businesses they have availed the services of, rate them, add photographs, tags, follow other people and their activities, etc. In this paper we examine and present the co-relation between a user's rating and the influence of the people, that the user follows, on the user for a particular business. The group of people that the user follows is commonly referred as friends of the user. We also analyze if a user can get influenced, if a business has a certain number of reviews already present or if the reviews have been written by elite reviewers (a reviewer who, according to Yelp, has contributed exceptionally in engaging the community in the form of consistency in writing reviews, as well as the quality of the reviews). Our analysis, through correlation and regression techniques, is able to prove that the user's rating remains unaffected by the number of people a user was friends with nor does the existing number of reviews and presence of elite reviewers helps in influencing a user. What shapes a user's rating is the overall experience, that the user had at the restaurant.  
\end{abstract}


\section{Introduction} \label{intro}
    The service based industry is growing rapidly in order to keep up with the pace of our lifestyle changes and serve our data intensive needs accordingly\footnote{source~(\url{https://www.plunkettresearch.com/industries/ecommerce-internet-technology-market-research/})}. The recent galore of Internet based service industry~\cite{bughin2011impact}, has presented immense possibilities and control over how, when and which service to choose from the plethora of existing ones~\cite{iqbal2003understanding,li1999impact}. This trend has sparked the onset of various such Internet based solutions, that enable the community to rate and/or review homogeneous genres of businesses. It empowers anyone having a particular requirement to be able to look anytime for the best possible service, which is available at a distance convenient to them. Such community driven Online platforms also allow businesses to serve their customers better~\cite{dwyer2015price,sharmaidentifying}, enabling a unique and transparent marketing channel for the businesses.
    
    Yelp is one such very popular, community driven effort which provides a platform to curate user reviews and ratings for local businesses, e.g. restaurants, salons, auto services, healthcare services etc. The quality of a business can be gauged by the average rating of the business, and the number and the connotation of the reviews tend to support the rating~\cite{huang2014improving}. Yelp allows its users to formulate social circles within each other by adding other users as friends. This enables the users to share with each other their feedback/experience regarding satisfaction with respect to a particular business. In such a social-driven rating scenario, it is often argued that the user business service preference and reviews are prejudiced to a certain degree derived by mutual friendship and  trust~\cite{yang2015predicting,iqbal2003understanding}. 
  
    We observe, from the yelp data, that the businesses, Food and Restaurant, contribute as the second largest category\footnote{\url{https://www.yelp.com/factsheet}} of all the business reviewed by Yelp users. We choose these two top categories of businesses for our study. In this paper, we examine: \textit{(i)} the impact of friends' experience on a user's rating, if any. For instance, whether or not, a user rating is indicative of the experience that his friends had; and \textit{(ii)} if there is any social influence of the number of reviewers or the presence of elite reviewers, on a user while rating a restaurant. For our study we employ the following approaches:
    \begin{itemize}
      \item Firstly, we carry out Pearson Correlation test on user rating and user friends' average rating. Along with that, using P value test, we also conduct a null hypothesis test.  \textit{and,}
      \item Secondly, we perform Linear Regression over user rating and user friends' average rating to gauge and visualize the relationship between the former and later.  Also, Linear Regression technique aids us with a few more statistical measurements to verify our claims.     
    \end{itemize}
    Both the approaches were also implemented for our second research objective, .i.e, understanding relationship between: user rating and number of reviewers; user rating and number of elite reviewers.
   
    The remainder of this paper is organised as follows: In \textit{Section}~\ref{relwork} we discuss the related work with respect to our current work. \textit{Section}~\ref{yelp} presents the dataset and tools used in our work, \textit{Section }~\ref{experiment} portrays the experimental setup that we did in the process. In \textit{Section}~\ref{socialimp} we discuss the findings from two main approaches employed and finally \textit{Section}~\ref{conc} concludes the paper by discussing the final outcome and the key statistics that helped us understand the results.  
    
\section{Related Work} \label{relwork}

 Yelp provides a rich dataset which has engendered a lot of research in the field of human behavior, social network analysis, recommendation techniques, etc. Our study has an essence of each of the above mentioned fields. In this section, we portray how our current work addresses the limitations of other related research efforts. Moreover, we discuss the idea of the contribution of various virtual or fictitious factors in influencing, primarily, the engagement of the community on the Yelp platform, rather than the user rating. 
    
    One of our major arguments that a user rating is predominantly, a reflection of his/her experience at the restaurant, is also proved with the help of Ridge regression technique~\cite{luca2011reviews}. This study is, however, based on the sentiments of the reviews written. The author zeroes down on a bag of 300 most popular and frequently occurring words in the reviews, and their present the maximum chances of influencing the ratings of a user. And these words describe the customers' experience, feel, facilities, ambiance, and quality of the neighborhood of the restaurant. 
    
    In one of the other studies~\cite{hajas2014analysis} which aligns to ours, in the sense, it looks at the overall sensitivity of the fluctuations in a restaurant rating and revenue as an attribute of 50 or more reviewers of that restaurant. It also studies the same fluctuations as an attribute of elite status of the reviewers. These are the people who have met certain criteria, decided by Yelp, who are lauded with such status so that they could give some more reliability or credibility to the reviews about the restaurant they rate. And hence, the author implies that in both of the cases above, a restaurant's overall perception and rating improves when the numbers of reviews are more than 50 and also that the rating is susceptible to people's opinion based on reviews by elite reviewers. 
    
    Whereas in our study, we touch a more humane aspect, in the sense that, how a user while rating a restaurant would not necessarily change his/her opinion altogether if he/she observes that, that restaurant had 50+ reviews or had reviews written by elite reviewers. In fact, user's opinions are formed as a result of several other factors and the actual experiences. Although the author's observation that the number of friends a reviewer might have on the website, may not necessarily impact his/her review about a restaurant, aligns with our observations as well. Although there is one difference, in the sense, that in this case, the author is weighting the overall rating by the number of friends each reviewer has. Whereas we do a case-by-case study, by finding out the pattern and behavior for every user rating for every restaurant versus the average ratings given by the user's friends to the same restaurant that the user rated and then we expand and generalize this observation to the entire dataset.
    
    Quite some work has been done in the field of user star rating prediction. In their study~\cite{hanpredicting}, the authors carry out user star rating prediction by analyzing the user and business features using similarity measurement algorithms. The study only takes into account, the user's past ratings in restaurants serving different categories of food. One of the other study, predicts user start rating based solely on the text of user generated reviews~\cite{fan2014predicting}. There also exists work, which discusses the contribution of social network in predicting user rating~\cite{yang2015predicting}. This study makes use of latent-factor model and then adds the user's friends' contribution to the business's ratings, in the model.
    
    While our study does not discuss the textual analysis of user reviews, the past ratings of the user or the latent-factor model in predicting user reviews, but it certainly does a lot more analysis than the above three studies. Our study discusses the various other aspects that determine how a user would rate a restaurant. From analysing the impact of friends experience, to the presence of elite reviewers and the number of overall reviewers, we also discuss upon the extrensic and intrinsic features of a restaurant like neighborhood quality, ambience and service quality among many others.

\section{Dataset and Tools} \label{yelp}

    \subsection{Yelp Dataset}
    
    We use the data from the famous Yelp Dataset Challenge\footnote{Yelp dataset~(\url{https://www.yelp.com/dataset_challenge})}. We have primarily used \textit{three} subsets of the Yelp dataset for our research, namely:  \textbf{business}, \textbf{reviews} and \textbf{user}. All these three subsets are in the form of individual JSON files. The dataset statistics are as follows:
        \begin{itemize}
    \item The \textbf{business} subset includes information of about 77K businesses and provides the following information, about a particular business: type, unique id, name, city, state, stars (which are rounded to half-stars), number of reviews, the category that the business falls in. 
    \item The \textbf{review} subset includes  information of about 2.2M reviews and tells us the following attributes of every review:the type of review, to which business id it belongs to, the user id who has written that review, number of stars given, date, votes, and the review text.  
    \item The \textbf{user} subset on the other hand gives information about 552K users and each user has: the unique id, the name, review counts, average stars of the user, friend ids, elite status in years, date of joining (in the form of Yelping since), compliments, fans, votes.  
    \end{itemize}
    
    \subsection{User ratings}
    In all of the above cases, stars or ratings, range from 1 to 5. And usually, the ratings are rounded off to half stars. So for example, a restaurant whose average rating is of 3.45, would have a rounded-off rating of 3.5. Also, a user is given an elite status for a particular year. He may or may not fall under the elite category year-on-year. And one of the ways in which a user usually makes a cut into the elite category, as per Yelp, is by contributing really good quality content (reviews, photos, tips, check-ins etc)  to the platform.
    
    \subsection{Tools}
    For our experiments, we used NumPy\footnote{NumPy library~(\url{http://www.numpy.org/})}, Scikit-learn\footnote{Scikit learn~(\url{ http://scikit-learn.org/stable/})}, Pandas\footnote{Pandas data analysis library~(\url{http://pandas.pydata.org/})}, Matplotlib\footnote{Matplotlib~(\url{http://matplotlib.org/})} libraries and code base is on \texttt{iPython notebook}. Moreover we specifically used the Pandas library for data analysis, Matplotlib for visualizations, Scikit-learn for carrying out regression analysis and NumPy for numerical calculations. We provide our complete source code publicly for research purposes here\footnote{\url{https://github.com/shubhanshu-gupta/yelp-data-challenge}}.

\section{Experimental setup} \label{experiment}

\subsection{Trimming the data}    

    As mentioned previously in Section~\ref{intro}, we use only those businesses that are mentioned in the 'Food' and 'Restaurant' category. Subsequently, the user subset containing information of about 552K users is trimmed down to approximately 250K by selecting only those users, who have atleast one (i.e. $>=1$) friend in their social circle. We divide the reviews file, having information of about 2.2M reviews, into multiple smaller segments for the ease of data handling.
    
\begin{figure}[htbp]
      \includegraphics[width=\linewidth]{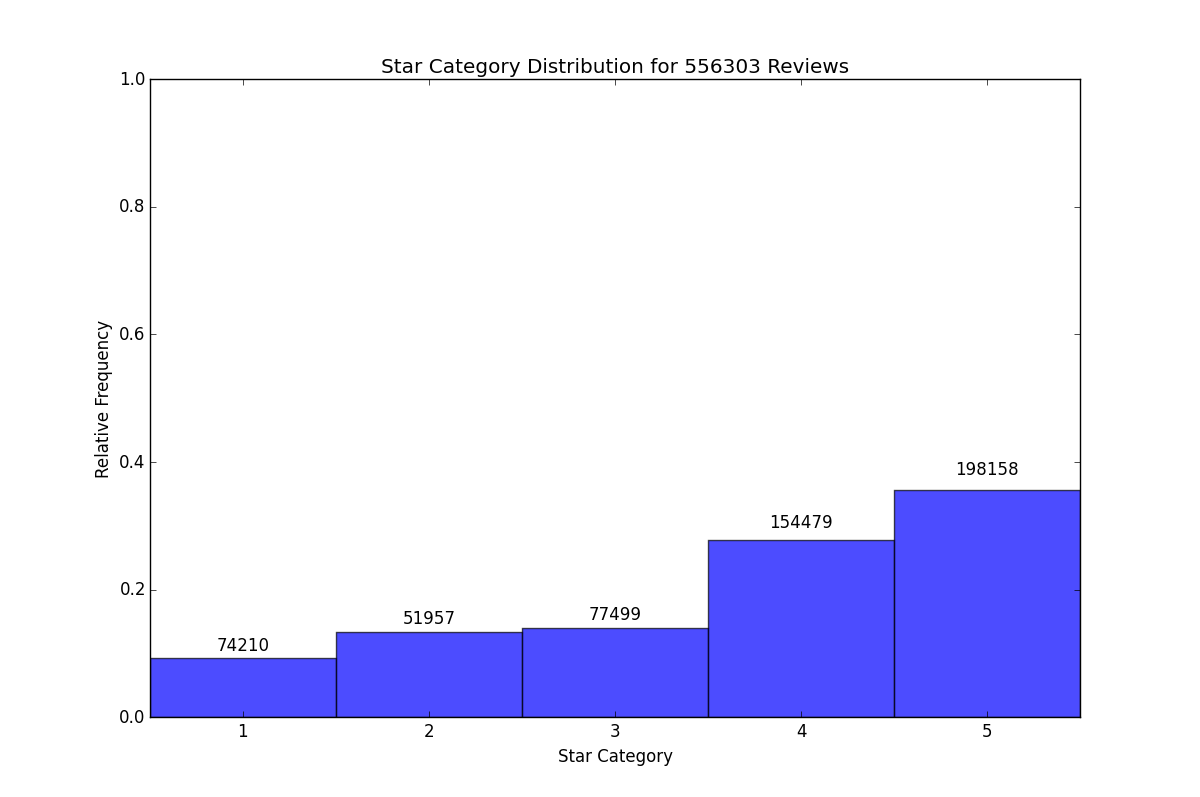}
      \caption{The star category distribution of the 1st quarter of reviews dataset}
      \label{fig:plot2}
    \end{figure}

    On performing a primitive data analysis on the first quarter of reviews subset, we observe that the star category distribution is  biased towards 4 and 5 stars, \textbf{ref. Figure~\ref{fig:plot2}}. These comprise of about 63\% reviews against the all the combined businesses having 1, 2, or 3 star ratings.
    
\subsection{Joining relevant data subsets}

    Subsequently, the first quarter of reviews (about 556K) is inner-joined with the key ids of users, so as to specifically get only those users who have written a review. This gives us a dataset of about 394K records. Also, this dataset is inner joined with the business subset on business ids (will be addressed as 'business merged dataset' in the rest of the paper), so as to specifically get only those businesses that are rated or reviewed by our select corpus of users. Having about 394K records, we select only those businesses which fall in the 'Restaurant' or 'Food' categories, thereby focusing only on 252K records. 

\subsection{Building data dictionaries}
\begin{figure}[htbp]
      \includegraphics[width=\linewidth]{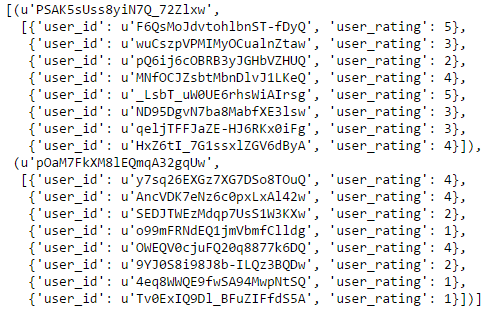}
      \caption{The Key (business id) - Value (list of dictionaries containing user ids and user ratings) pairs of second dictionary}
      \label{fig:plot4}
    \end{figure}

    In order to handle this large amount of data (i.e. more than 252K rows) efficiently, we build hash-maps or ordered dictionaries. The first dictionary that we build is of user (key) and list of their friends (value). We traverse two datasets: the user subset; the 'business merged dataset'. The former is traversed in order to fetch users and their friends, thereby leading our way to the first dictionary. And the latter, is grouped by business ids which gives us the lists of user ids and user ratings, as values, for every business id as the key - and this forms our second dictionary. This dictionary looks something like \textbf{ref. Figure~\ref{fig:plot4}}
    
\subsection{Deriving experiment-ready 'final dataset'}    

    We traverse the business merged dataset along with using the ancillary dictionaries to build a dataset of derived columns. So for every user and business, the dataset finally comprises of derived columns including:

    \begin{itemize}
      \item Number of friends who reviewed the same business as the user did  
      \item Average friends rating  
      \item Number of elite reviewers  
      \item Number of friends who were elite 
      \item Total number of reviewers etc.
    \end{itemize}
    
    This results in a dataset of about 90K records which contains the information of about 4K unique restaurants. Hence, each record in our final dataset consists of a business id, the user id of the reviewer, business average rating, total friends this user has, average rating of these friends of user, number of friends who rated this business id, total number of reviewers who rated this business, \& number of elite reviewers. We, from this point onwards, address this dataset as 'final dataset'.

    \begin{figure}[htbp]
      \includegraphics[width=\linewidth]{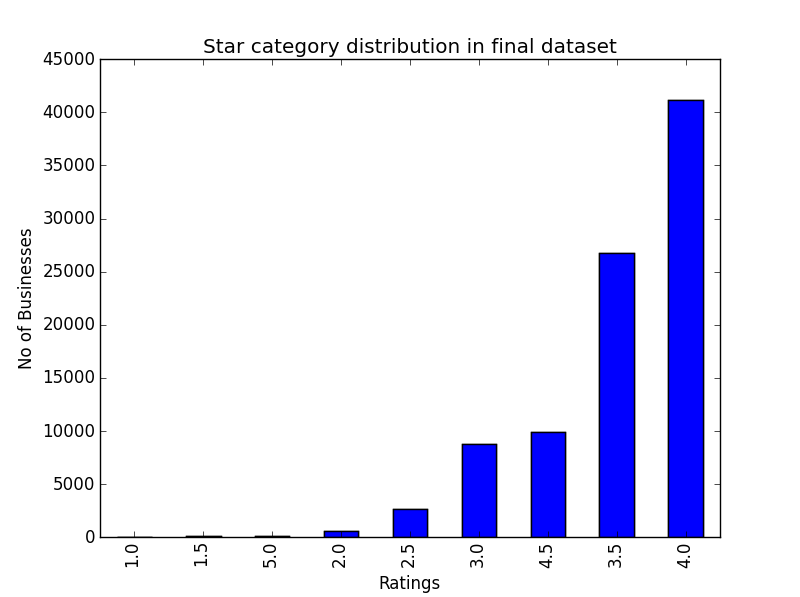}
      \caption{Star category distribution in the 'final dataset'}
      \label{fig:plot3}
    \end{figure}
    
    It can be visualized that the star rating distribution of the derived 'final dataset', has a little different pattern from its earlier counterpart \textbf{ref. Figure~\ref{fig:plot3}}. Here, about 90K users give ratings which are more skewed towards 4.0 and 3.5 stars for about 4K unique businesses. They comprise of about 75\% of the total records whereas all the others combined, comprise of just 25\% of the total records.

\section{Social impact analysis on yelp ratings} \label{socialimp}
    In this section we analyze, whether the ratings given on Yelp have a social edge or not. By social edge, we try to understand how a user rating for a business could be influenced by his/her friends' opinion for that same business.

    \subsection{Correlation analysis}
    \textit{Correlation} basically measures the linear correlation between two variables, user rating and friends' average rating, in our case. The correlation value varies between -1 to +1, wherein 0 indicates that there is no relationship between the two variables. And the perfect correlation of -1 or +1 implies that both the variables are exactly linearly related. 
    
    The correlation value for the friends' average rating column and the user rating column in our 'final dataset' obtained is 0.3219. This, although, is on a positive side, but is very weak to imply any strong linear relationship between user rating and friends' average rating. Our null hypothesis that, there exists a linear relationship between user rating and user friends average ratings is rejected by the meager p-value. The p-value is recorded below a bare minimum threshold which made it indistinguishable and therefore, came up to be 0.0. Hence, our alternative hypothesis, that there is an invalidated relationship between user and friends average ratings, gets proved. 
    
    \begin{table}[]
    \centering
    \caption{The matrix of correlation between User and Friends' Average Ratings}
    \label{my-label1}
    \begin{tabular}{|c|c|c|}
    \hline
     & \textbf{\begin{tabular}[c]{@{}c@{}}Friend's average \\ rating\end{tabular}} & \textbf{\begin{tabular}[c]{@{}c@{}}User's average\\ rating\end{tabular}} \\ \hline
    \textbf{\begin{tabular}[c]{@{}c@{}}Friend's average \\ rating\end{tabular}} & 1.0000                                                                      & 0.3219                                                                   \\ \hline
    \textbf{\begin{tabular}[c]{@{}c@{}}User's average\\ rating\end{tabular}}    & 0.3219                                                                      & 1.0000                                                                   \\ \hline
    \end{tabular}
    \end{table}
    
    \begin{table}[]

    \centering
    \caption{Results of Pearson Correlation Test}
    \label{my-label2}
    
    \begin{tabular}{|l|l|l|}
    \hline
    \textbf{R value} & \textbf{P value} & \textbf{Stderr} \\ \hline
    0.3219           & 0.0              & 0.0040          \\ \hline
    \end{tabular}
    \end{table}
    
    \subsection{Regression analysis}
    In order to visualize and validate the above hypothesis, we further decided to do a \textit{regression} analysis on our 'final dataset'. We divided the 'final dataset' into 70-30 ratio of training and test data, respectively. The dependent variable was friends’ average rating column and the independent variable was user rating column. The coefficient value was 0.225, and an intercept value was 2.984. And with the R square value of just 0.088, it means that more than 90\% of the variation is residing in the residuals. But, to be sure about how truthful the value of R square was, in indicating how close our predicted data is to the fitted regression line, we plot a scatterplot to get a sense of how well the data seems to fit. 
    
    \begin{figure}[htbp]
      \includegraphics[width=\linewidth]{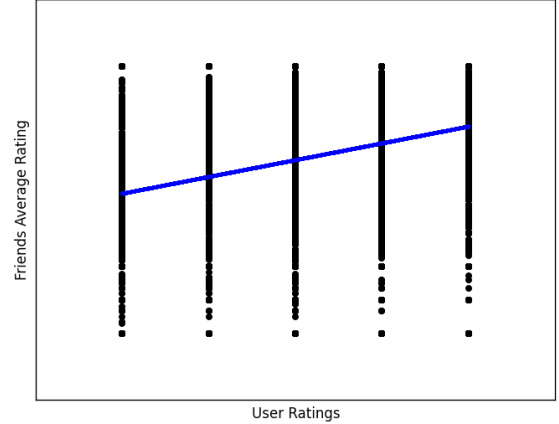}
      \caption{The combination of the scatter plot (black) between User and Friends’ Average Ratings and the plot (blue) of the linear relationship between the two.}
      \label{fig:plot1}
    \end{figure}
    
    The plot \textbf{ref. Figure~\ref{fig:plot1}} is a visual evidence of the fact that as the user rating increases or decreases, friends average rating does not follow the same trend, and instead it proves that every person rates based on his/her own experience with the business. The user rating and the predicted user friends' average ratings are represented by the blue line, and the scatter plot of friends' average ratings and user ratings are represented by the black dots. The complete vertical black dots indicate that for a user rating, for ex- 2, the friends’ average could vary from 1 to 5, extensively. And same holds true for higher user ratings of 4 and 5. This certainly affirms that the correlation between Friends' average ratings and user ratings is negligible, which gives us the confidence to derive the conclusion that for a user who rated highly (or low) for a business, the friends in his/her social group may not have the same experience with the same business. 
    
    Furthermore, we also check for the presence of any correlation between the user ratings, a business receives and the number of reviewers it has. We normalize the dataset, comprising of number of reviewers column as the dependent variable and user rating column as the independent variable. Followed the same approach as of regression analysis and found that with a meager R square value of 0.009, almost all the variations reside in residuals. And to verify further, we also ran a Pearson test from the Scipy library and obtain the R value to be 0.09, P value was in negative power of e, and std err was 0.005. Adding up all of the above results, it's pretty evident that a user while rating a business does not get influenced by the number of reviews that business has. 
    
    Finally, we also run similar experiments to understand whether there is any influence of the presence of reviews by elite users on a general user rating. And, since in all of the above experiments we mostly came up with a similar conclusion that a user rates based on his/her experience solely, it came as a no surprise that here also, a user does not get influenced while rating for a business which has reviews by users who are of elite status.

    \subsection{Results}
    Our study of the level of satisfaction or dissatisfaction for a user and his/her experience at a restaurant, as a reflection of his/her social circle on Yelp, number of reviews and the reputation of other reviewers is \textit{validated} by a number of statistical figures.  We'll go through them briskly to understand how they shaped and affirmed our study. 
    
\begin{figure}[htbp]
      \includegraphics[width=\linewidth]{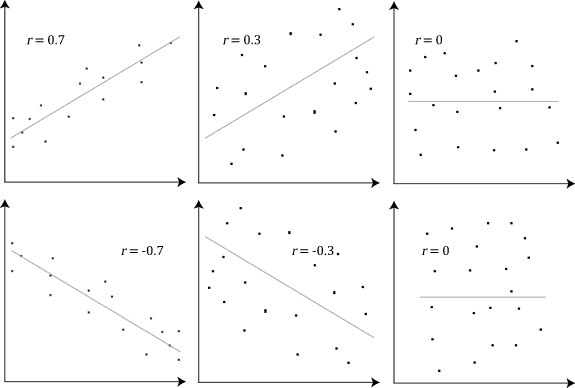}
      \caption{The figures seemingly portray the correlation between the data points, with the appropriate R values.}
      \label{fig:plot5}
    \end{figure}

    Pearson correlation coefficient or the R value\footnote{Pearson coefficient- \url{http://goo.gl/hSHI4I}}, measures how strongly two variables are related linearly two each other. The coefficient which ranges between +1 to -1 (where both the extremes depict best fit), indicate the level of association between those two variables. A simple figure to understand the R value can be understood from \textbf{ref. Figure~\ref{fig:plot5}}.
    
    The second thing we elaborate is P value\footnote{P value~(\url{http://goo.gl/xq4jmJ})}. It helps in determining the significance of the results when a hypothesis is tested. Our null hypothesis was that there existed a linear relationship between user rating and friends' average ratings. The value of P-value was so meager which weighed our evidences (the data points in our final dataset) so correctly that we were not only assured of our methodology by which we derived our final dataset but also it helped us reject our null hypothesis. Usually this happens when the P value is less than equal to 0.05. 
    
    Next, we elaborate our observations from linear regression analysis. It's a predictive modeling methodology which draws out a fine relationship (both, if there is some or none at all) between a dependent variable and an independent variable\footnote{Regression Techniques~(\url{http://goo.gl/GqXEzh})}. Like, Pearson correlation coefficient, it helps in indicating the strength of relationship between the two variables and thereby, predicts the future patterns on the similar lines. Hence, while performing the regression analysis (we performed linear regression), we are served with multiple statistical figures like coefficient, slope, intercept etc. The plot of our linear regression depicts and the r square value measures the closeness of our data to the fitted (blue) regression line. R square\footnote{Interpreting R square~(\url{http://goo.gl/xj5ceu})} which can also be understood by the ratio of explained variation to the total variation and it ranges between 0 and 1. In general, when the R square value is high, the model represents a good fit our dataset. Since, our R square value was so low, that also became a key indicator of the lack of variance accounted by our regression model which made our data points spread all across the fitted regression line. This was also supported by our residual plots which, as a result housed more than 90\% of our variations.
    
\section{Conclusions} \label{conc}
In this paper, we examined and evaluated the presence or more aptly, the absence of the contribution of factors that define how a user would rate his/her experience with a business. We, in particular, identify whether a user rating is more or less influenced, by modest extrinsic factors surrounding a restaurant and several intrinsic factors that lie within a restaurant~\cite{hajas2014analysis,dwyer2015price}. For instance, the level of satisfaction or dissatisfaction reflected by a user in their rating for a business could be driven by: \textit{(a)} service (slow, good, bad, etc.), \textit{(b)} food quality, \textit{(c)} quantity or ambiance with respect to the price of the meal, \textit{(d)} overall experience (and hygiene) at the locality and also \textit{(e)} the popularity of the neighborhood in which restaurant is situated~\cite{sharmaidentifying}. 

On the other hand, there is no direct relationship between how a group of people, whom we simply follow on Yelp, might have rated a restaurant. In reality we may not even know these people, and neither their tastes, or preferences. And that's why, a user's social circle defined within Yelp's enclosure does not contribute much to a user's rating - which has been corroborated by our research findings. There is also no relationship between how a user would rate a restaurant and how many reviews that restaurant already holds or how many of the reviewers of the restaurant are/have been elite as per Yelp. These factors are, probably, best to substantiate the overall picture of a restaurant that Yelp portrays with the help of aggregated reviews, ratings, photographs, etc. They do help to sort and filter out a restaurant based on an individual's or a group's mood, taste and preference at that point of time, but none of these factors determine how the individual or a group would feel or experience, while they are actually at the restaurant, having their meal. 

\bibliographystyle{abbrv}
\bibliography{sig} 
\end{document}